\documentclass{rQUF2e}

\usepackage{epstopdf}
\usepackage{subfigure}
\usepackage{graphicx}
\usepackage{float}

\theoremstyle{plain}

\theoremstyle{definition}

\theoremstyle{remark}

\begin{document}
\renewcommand*{\refname}{}
\bibliographystyle{rQUF}


\title{Election predictions are arbitrage-free: response to Taleb}

\author{AUBREY CLAYTON$^{\ast}$$\dag$\thanks{$^\ast$Corresponding author.
Email: aubreyiclayton@gmail.com
\newline
\dag The views expressed in this publication are wholly those of the author. They do not necessarily represent the views of the author's employer or any of its affiliates; and accordingly, such employer and its affiliates expressly disclaim all responsibility for the content and information contained herein.
}\\
\received{} }

\maketitle

\begin{abstract}
\cite{taleb2018} claimed a novel approach to evaluating the quality of probabilistic election forecasts via no-arbitrage pricing techniques and argued that popular forecasts of the 2016 U.S. Presidential election had violated arbitrage boundaries. We show that under mild assumptions all such political forecasts are arbitrage-free and that the heuristic that Taleb's argument was based on is false.
\end{abstract}


\begin{classcode}C11; C53
\end{classcode}

\section{Introduction}

Forecasts of a candidate's probability of winning an election have become a staple of political journalism over the last decade, with the most prominent being those reported on the website FiveThirtyEight, for example in the 2016 U.S. Presidential election; see \cite{538}. \cite{taleb2018} was motivated to critique the quality of election forecasts by the apparent instability of these probabilities through time. FiveThirtyEight reported a probability of Clinton winning the 2016 election that ranged from 55 to 85 percent over the final five months. According to Taleb, this was indicative of `stark errors' and `severe violations' of standard results in quantitative finance, specifically no-arbitrage option pricing. Imposing a no-arbitrage constraint on election forecast probabilities, as binary options written on the candidate's vote share, would therefore, according to Taleb, eliminate such instabilities. 

In this paper we show that Taleb's argument was mistaken. First, one of the `standard results' of quantitative finance that his election forecast assessments rely on is false, as we demonstrate with a simple counterexample. Next, we show that binary option prices can easily exhibit as much or more variability through time as the 2016 Presidential election forecasts without violating any no-arbitrage constraints. We argue that, under mild assumptions concerning the information available to a forecaster, all such election forecasts are arbitrage-free. Finally, we comment on a problem regarding an ill-defined variable in Taleb's election model and what this implies for the general applicability of option pricing methods to election forecasting.

\section{Summary of Taleb's argument}

We being by summarizing Taleb's argument, which consists of five main steps. We also take this opportunity to fix some notation and definitions:

\begin{enumerate}

\item He claims that the forecast probability for an event, such as a candidate winning an election, should be considered as the price of a binary option taking the terminal value 1 if the event happens and 0 otherwise. This can be thought of a derivative security on an underlying `asset' variable $Y_t$, e.g., the candidate's number of supporters among the population of voters at a given time $t$, with the election outcome determined by whether $Y_T$ exceeds some threshold $l$ at terminal time $T$. Arbitrage pricing theory will then dictate limits on the behavior of these prices/probabilities over time. 

\item He gives a model for the stochastic process governing the underlying asset $Y_t$. To produce a $Y_t$ that is bounded in some range $[L, H]$, Taleb assumes $Y_t = S(X_t)$ for a `shadow process' $X_t$ satisfying a tractable stochastic differential equation, where $S$ is a function taking the real line to a finite interval. He makes the choice: 

$$dX_t = \sigma^2 X_t dt + \sigma dW_t$$

and 

$$S(x) = \frac{1}{2} +  \frac{1}{2} erf(x)$$

where $erf(x) = \frac{2}{\sqrt{\pi}} \int_0 ^ x \exp(-t^2) dt$. 

\noindent Note that 

$$\frac{1}{2} S''(x) + x S'(x) = 0$$

With that choice in place, It\^{o}'s formula implies $Y_t$ satisfies

$$dY_t = \left( \frac{1}{2} \sigma^2 S''(X_t) + \sigma^2 X_t  S'(X_t) \right) dt + \sigma S'(X_t) dW_t $$
$$= 0 \cdot dt + \sigma S'(X_t) dW_t$$

That is, the process $Y_t$ is a bounded martingale. 

\item Since the process $X_t$ is the familiar Bachelier-style model for an asset price, the price of a binary option of $X_t$ has a known formula. Arguing that this price is equal to the price one would obtain for the binary option on $Y_t$ with the corresponding threshold value, this gives the election forecast probability. Substituting $x = S^{-1}(y)$ into the option pricing formula for $X$ gives the price in terms of the currently observed value of $Y_t$ at a given time. Since $Y_t$ is a martingale, Taleb argues that the stochastic equation for $Y_t$ represents the dynamics under the risk-neutral measure, and therefore the price obtained for the binary option represents the probability of $Y_T$ exceeding the threshold, that is, the probability of the given candidate winning the election. Since the forecast probabilities are derived as option prices, under this measure they are also martingales.

\item A heuristic from option pricing suggests that the greater the volatility of the underlying asset, the closer the binary option price should be to $0.5$ and the lower the volatility of the option price should be over time. A numerical example for the forecast probability of Trump winning the 2016 Presidential election shows that Taleb's model gives prices close to $0.5$ until just before election day, at which point it jumps to near $1.0$, in agreement with the heuristic argument. Observing electoral forecasts such as those made by FiveThirtyEight show forecast probabilities deviating significantly from $0.5$ and fluctuating widely, even at a date far in advance of the election. He claims this means the FiveThirtyEight forecasts must have `violate[d] arbitrage boundaries.'

\item Taleb then compares his martingale-pricing approach with an argument due to de Finetti \citep{definetti2009} that forecast probabilities be evaluated by means of the Brier score 

$$\frac{1}{n} \sum_{i=1}^n (\mathbf{1_i} - p_i)^2$$

where $p_1, \ldots, p_n$ are prices of bets laid on the outcomes of $n$ independent events, and $\mathbf{1_i}$ is the indicator function taking the value 1 if the event occurred and 0 otherwise. In order to minimize losses according to this score, an agent will place bets in agreement with their probabilities for the events. He claims the martingale-pricing technique gives a continuous time analogue of the same idea.

\end{enumerate}

\section{Criticism}

\subsection{The behavior of option prices with respect to volatility}
Taleb's heuristic regarding the behavior of binary option prices as the volatility of the underlying increases is only partially true. He claims that `[a] standard result in quantitative finance is that when the volatility of the underlying security increases, arbitrage pressures push the corresponding binary option to trade closer to 50\%, and become less variable over the remaining time to expiration.' It is the case that for a given `spot-price' $Y_t$ and a given threshold value $l$, as the volatility $\sigma$ in the stochastic model for $Y_t$ is increased, under general conditions the binary option price/probability of exceeding the threshold should converge to $0.5$. For example, if the stochastic process for $Y_t$ were a simple Brownian motion with drift

$$dY_t = \mu dt + \sigma dW_t$$

\noindent the conditional distribution for $Y_T$ given $Y_t$ would be $N(Y_t,\sigma^2 (T-t))$ and so the binary option price for strike $l$ at time $t$ would be

$$B(t,T) = 1 - \Phi \left(\frac{l - Y_t - \mu(T-t)}{\sigma \sqrt{T-t}} \right)$$ 

\noindent where $\Phi$ is the standard normal cdf. This converges to $0.5$ as $\sigma$ goes to $\infty$. Other processes depending on a volatility parameter will show the same behavior.

However, the prices $B(t,T)$ one actually \emph{observed} over time would not stabilize around $0.5$, because the higher volatility for the process $Y$ would also result in a wider distribution for $Y_t$ at any given $t$. Taking the same example above with $Y_0 = 0$, we have an unconditional cdf for $B(t,T)$ given by

\begin{equation*}
\begin{aligned}
P[ B(t,T) < x] = P \left[ \Phi \left(\frac{l - Y_t - \mu(T-t)}{\sigma \sqrt{T-t}} \right) > 1 - x \right] \\
= \Phi \left( \frac{l - \mu (T-t) - \sigma \sqrt{T-t} \Phi^{-1}(1-x) - \mu t}{\sigma \sqrt{t}} \right)  \\
= \Phi \left( \frac{l - \mu T}{\sigma \sqrt{t}} - \sqrt{\frac{T}{t} - 1} \Phi^{-1}(1-x) \right)
\end{aligned}
\end{equation*}

So, as $\sigma \rightarrow \infty$, we end up with a distribution for $B(t,T)$ that is stable but certainly not a point mass at $0.5$. In particular, at the midway point $t = T/2$, if we have set our threshold $l$ to be equal to the projected mean of $Y_T$, $l =\mu T$, then we will have 

$$P[B(t,T) < x] = \Phi \left( - \Phi^{-1}(1-x) \right) = 1 - \Phi(\Phi^{-1}(1-x)) = x$$

\noindent and so $B(t,T)$ has the \emph{uniform} distribution on $[0,1]$ regardless of $\sigma$. 

This counterexample shows that Taleb's second claim, that as volatility increases binary option prices `become less variable over the remaining time to expiration,' is false. Instead, the variability of the paths of the option price may be completely independent of the volatility in the underlying. This has profound implications on his criticism of election forecasts. For example, assuming a simple random-walk model for a given candidate's vote share, if the process stays close to the value $50\%$ we may observe the win-probability fluctuate widely through time despite arbitrarily small fluctuations in the polls. The figures below show one such simulated path over 1000 days where we have set the size of daily poll movements to 0.1\%. We could just have easily chosen 0.000001\% or any small number. 

\begin{figure}[H]
\centering
\includegraphics[width=.8\linewidth]{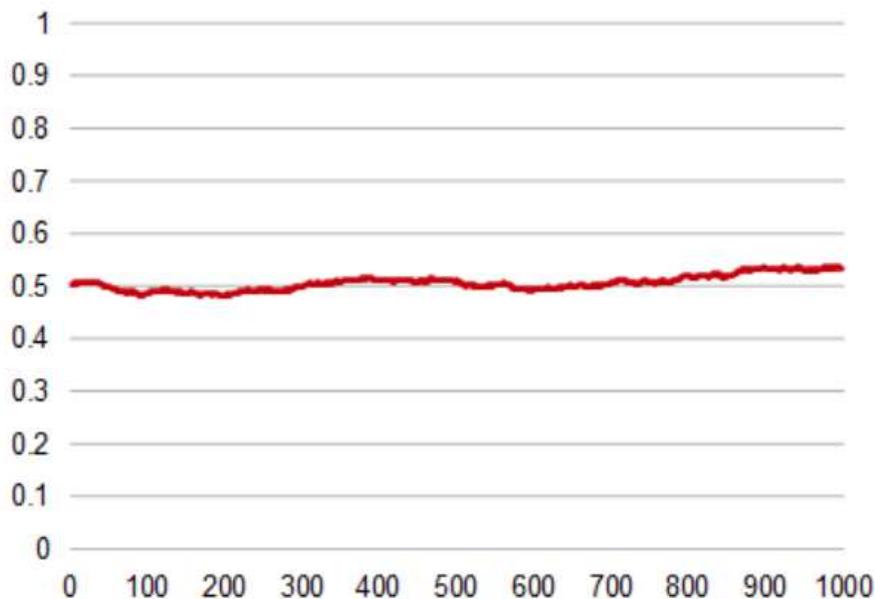}
\caption{Simulated candidate vote share for a random-walk process; step = 0.01\%}
\label{fig:randomwalkpolls}
\end{figure}

\begin{figure}[H]
\centering
\includegraphics[width=.8\linewidth]{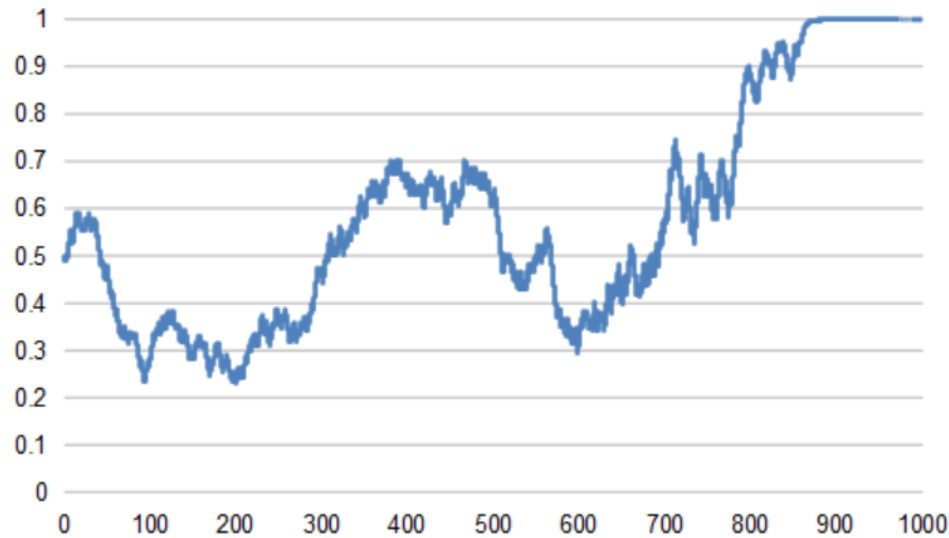}
\caption{Simulated candidate win-probability for a random-walk process; step = 0.01\%}
\label{fig:randomwalkwinprob}
\end{figure}

\noindent The size of these fluctuations in poll numbers and forecast win-probability are roughly consistent with what was observed in 2016. 

In fact, even in Taleb's model, if the same \emph{s} parameter, representing annualized volatility in the polls, is used to project the polls as well as construct the forecasts, the paths do not stabilize at $0.5$ for high volatilities. For example, the figure below shows one such path with $s = 100\%$:

\begin{figure}[H]
\centering
\includegraphics[width=.8\linewidth]{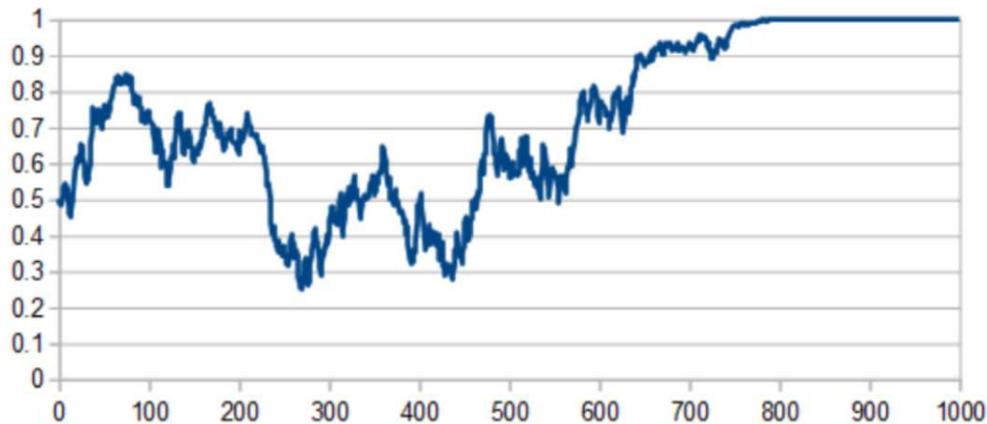}
\caption{Simulated candidate win-probability for Taleb's pricing model; s = 100\%}
\label{fig:talebwinprob}
\end{figure}

Intuitively speaking, the greater we assume the implied volatility for the pricing model $Y_t$ to be, the greater we must also assume the realized volatility to be; the former pulls option prices in toward $0.5$ for any given spot-price while the latter widens the distribution of the actual spot-price we observe. Since the same volatility we observe is priced into the option, the two effects more-or-less exactly cancel, leaving the paths of forecast probabilities/prices unchanged. For Taleb's demonstrated stable probabilities to manifest, we would need to consistently use a higher implied volatility than what has been realized, an `excess volatility' assumption that he does not explicitly state or attempt to justify.

\subsection{Election predictions are arbitrage-free}

Framing election forecasting as a problem of option pricing adds needless complication and introduces the possibility of confusion. Taleb's argument relies on an analogy to the problem: given a model for the dynamics of an asset price under real-world probabilities $\mathbb{P}$, price an option on the asset. The Fundamental Theorem of Asset Pricing implies that in a complete market with no arbitrage there is an equivalent probability measure $\mathbb{Q}$ under which all assets earn the risk-free rate $r$ on average, including the option. Setting $r=0$ would make all assets martingales under this measure. Thus, the value at any given time for any derivative security is simply the average of its possible values at any future time; for a binary option this implies the price is equal to the probability of occurrence for the payoff event. 

However, the task facing a forecaster is not one of pricing an option but of assigning a real probability. Suppose this is done over the filtration $(F_t)$ of information available to the forecaster at any given time. Granting Taleb's assertion that the probabilities should be thought of as prices, and that these prices should be martingales under the \emph{real-world measure}, we still needn't bother with the artifice of defining a stochastic model for $Y_t$ under which it is a martingale and pricing an option written on $Y_t$. Instead, we can achieve exactly the dynamic Taleb desires by considering any stochastic process at all for $Y_t$ and then quoting the probability 

$$B(t,T) = \mathbf{P}[Y_T > l | F_t]$$

These probabilities/prices always produce an $(F_t)-$martingale by the tower property of conditional expectation:

$$E[ B(t,T) | F_s ] = E[ E[ \mathbf{1}_{Y_T > l}] | F_t] | F_s] = E[ \mathbf{1}_{Y_T > l}] | F_s] = B(s,T)$$

\noindent for any times $s < t$. 

Thus, Taleb's construction of a martingale $Y_t$ is sufficient but not necessary for the forecast probabilities to be martingales, which makes his shadow process $X_t$ doubly unnecessary. Taleb's construction requires a bounded martingale that can be written as some monotonic function of a process for which the option prices were possible to compute, but he could have skipped that entirely. 

Different forecasters may disagree in their assessments, and the results may be profitable for one or the other. Whether the forecaster's probabilities allow for profitable investments over time will depend on the judgments of the forecaster and the investor. However, since the forecaster's conditional probabilities are automatically martingales with respect to their own filtration, there is no possibility for arbitrage unless the two assigned probability measures are not equivalent. In order to exhibit arbitrage, then, Taleb would need to show an event a forecaster assigned nonzero probability to was actually impossible, or conversely, a forecaster claimed an actually possible event had probability zero.

\subsection{The definition of the underlying}
Finally, Taleb's analysis suffers from a lack of precision about the meaning of the `underlying' asset $Y_t$, from which the election results are meant to be determined. He desires it both to be the vote count/share $Y_T$ at the terminal date (or number of electoral votes for determining a presidential election outcome) and yet also to be measurable by some means at times $t < T$ as `an intermediate realization of the process at t.' This would be possible if the opinion of every potential voter in the U.S. were known at every given time, in which case the volatility of $Y_t$ would correspond to changes of \emph{opinion}. 

However, the problem for the forecaster is that the intentions of every person are not known and must be estimated from polls/samples. Thus, the uncertainty facing a political forecaster such as FiveThirtyEight is that these samples themselves have uncertainty, as do the proportions of people who actually vote in the election. Taleb later defines $Y_t$ to be `the observed \emph{estimated} proportion of votes' (emphasis ours). If we treat $Y_t$ as the estimated proportions of voters, as given by the samples, we would find it inadequate to determine the election of the outcome; elections are decided by the actual vote counts, not a final random sample of voters on election day. This means that the process $Y_t$ cannot possibly be both an underlying asset for the derivative and adapted to the filtration of information available to the forecaster.

\section{Conclusion}

Taleb's criticism of popular forecast probabilities, specifically the election forecasts of FiveThirtyEight, was inspired by a judgment that they tend to fluctuate too much given a reasonable amount of uncertainty in the future changes in popular opinion. He attempted to cast this as a violation of the principles of option pricing and to produce an alternative model by constructing a bounded martingale with known binary option prices exhibiting the dynamics he sought. However, the situation facing the forecaster is one of assigning probabilities, not prices, and this can be done in the real-world measure without reference to an underlying martingale asset. As long as these probabilites are updated consistenly according to rules of conditional probability, they will automatically be martingales. If the forecaster's probability measure is equivalent to an investor's, then according to the Fundamental Theorem of Asset Pricing arbitrage is impossible even if the forecasts are treated as prices. There may be profitable, but risky, market opportunities if investors have different probability assessments than the forecasters, but this is simply an argument that the investors' forecasts are better. 

Taleb's heuristic of greater volatility/uncertainty giving binary option prices that stabilize around $0.5$ suggested that these forecasts had violated arbitrage boundaries, presumably, by understating the amount of uncertainty. However, that heuristic is neither true nor relevant. The paths of the conditional probabilities of a stochastic process terminating above some threshold level do not necessarily converge to constant lines at $0.5$ as volatility increases. It's possible to produce counterexamples where the distribution of prices at some intermediate point may be uniform on $[0,1]$, independent of the volatility of the underlying. This implies the observed variability in election forecasts is not necessarily indicative of any errors at all. The forecast probabilities may have fluctuated because they priced in only as much volatility as had been observed in the polls, a reasonable assumption that Taleb does not argue against. In any event, this is not a question of arbitrage.

Finally, the lack of clarity as to what the underlying asset $Y_t$ represents in Taleb's model speaks to a misunderstanding of the kinds of uncertainty facing the forecaster. In quantitative finance, an asset with an observable price is given a stochastic model, and fluctuations in that price determine the uncertainty of the terminal payoffs of various derivatives. In election forecasting, the ``price'' itself is unknown and unknowable. Thus, the whole framework of derivative pricing is arguably inapplicable here, which is no great loss since there was no need for it in the first place.

\section{References}

\bibliography{talebreferences}

\end{document}